\documentclass[twocolumn,trackchanges]{aastex701}
\usepackage{amsmath}
\usepackage{graphicx}
\usepackage{dcolumn}
\usepackage{bm}
\usepackage{xcolor}

\begin{document}

\title{Particle acceleration and pitch angle evolution in relativistic turbulence}

\correspondingauthor{Daniel Humphrey}
\email{dahumphrey@wisc.edu}

\author{Daniel Humphrey}
\affiliation{Department of Physics, University of Wisconsin at Madison, Madison, Wisconsin 53706, USA}
\email{}

\author{Cristian Vega}
\affiliation{Department of Physics, University of Wisconsin at Madison, Madison, Wisconsin 53706, USA}
\email{}

\author[orcid=0000-0001-6252-5169]{Stanislav Boldyrev}
\affiliation{Department of Physics, University of Wisconsin at Madison, Madison, Wisconsin 53706, USA}
\affiliation{Center for Space Plasma Physics, Space Science Institute, Boulder, Colorado 80301, USA}
\email{}

\author[orcid=0000-0003-1745-7587]{Vadim Roytershteyn}
\affiliation{Center for Space Plasma Physics, Space Science Institute, Boulder, Colorado 80301, USA}
\email{}

\begin{abstract}
Synchrotron radiation detected from relativistic astrophysical objects such as pulsar-wind nebulae and {jets from active galactic nuclei} depends on the magnetic fields and the distribution functions of energetic electrons in these systems. Relativistic magnetically dominated turbulence has been recognized as an efficient mechanism for structure formation and non-thermal particle acceleration in these environments. Recent numerical simulations of relativistic turbulence have provided insights into the energy distribution functions of accelerated electrons. Much less is currently understood about their {pitch angle distributions}, which are crucial for accurately interpreting the spectra of synchrotron radiation. {We perform a detailed case study of} the pitch angle distributions formed during the process of turbulent acceleration {for $B_0/\delta B_0 = 10$ and $\tilde{\sigma}_0 \sim 40$, where $B_0$ is the uniform component of the magnetic field, $\delta B_0$ is the fluctuating component, and $\tilde{\sigma}_0$ is the plasma magnetization based on the magnetic fluctuations.  We find that even minimal numerical noise can cause substantial pitch angle scattering, but we demonstrate techniques for overcoming the numerical challenges associated with the evolution of very small pitch angles. Our numerical results are consistent with the phenomenological model found in \cite[][]{vega2024b,vega2025}.}

\end{abstract}

\keywords{\uat{Galaxies}{573} --- \uat{High Energy astrophysics}{739}}


\section{Introduction} 
Large-scale plasma flows existing in various important astrophysical systems, including the outer solar corona, solar wind, interstellar medium, pulsar wind nebulae, and {jets from active galactic nuclei (AGN)}, are nearly collisionless, meaning that the typical nonlinear interactions and processes of particle energization happen on time scales significantly shorter than those associated with collisional relaxation. In these environments, the electron energy distribution functions can deviate significantly from equilibrium shapes. {Processes that accelerate electrons to ultrarelativistic energies generate non-thermal tails in the energy spectra, which in turn can cause} the frequency spectra of the resulting synchrotron radiation to become harder \cite[e.g.,][]{atoyan1996,meyer2010,abdo2011a,abdo2011b}. Generally, the intensity and spectral characteristics of synchrotron radiation depend on several factors: the {electron energy distribution}, the {strength} of the {local} magnetic field, and the electron pitch angle, defined as the angle between the electron momentum vector and the magnetic field direction \cite[e.g.,][]{pacholczyk1970}. 

The pitch angle is often considered unimportant since it is assumed to be uncorrelated with the electron energy, and results are averaged over {an isotropic} pitch angle distribution. This assumption is reasonable if some mechanisms of pitch angle scattering are present. For example, interactions between electrons and plasma waves or turbulence can lead to pitch angle {isotropization}. Recently, interest has been drawn to processes of particle acceleration by turbulence in which the pitch angle of accelerated particles is not random but rather correlated with the particle energy \cite[e.g.,][]{sobacchi2019,sobacchi2020,comisso2020,nattila2022,sobacchi2023,vega2024,vega2025}. 
{The assumptions about the pitch angle of energetic electrons can significantly influence the analysis of radiation spectra from astrophysical objects. For example, as discussed in \cite[][]{tavecchio2016,sobacchi2019,comisso2020,sobacchi2023}, different assumptions regarding the pitch angle distributions of radiating electrons in blazars—whether they are isotropic or highly collimated along the background magnetic field—can lead to notably different estimates of magnetic energy and electron cooling times in these systems.}

Strongly anisotropic electron distributions may arise when the acceleration process preserves the particle's magnetic moment, the first adiabatic invariant. This occurs when the electron gyroradius is significantly smaller than the inner scale of turbulent fluctuations, and the gyrofrequency is considerably larger than the frequency {corresponding to the characteristic dynamical lifetime of} the smallest and fastest turbulent eddies \cite[e.g.,][]{vega2024,vega2025}. Such a situation arises  when electrons are accelerated by turbulence in the presence of a relatively strong guiding magnetic field, that is, when the turbulent magnetic fluctuations, $\delta B_0$,  are weaker than the uniform component, $B_0$, associated with the magnetic flux permeating the system. {A} strong guide field ensures that the gyroradius remains small, which imposes a restriction on the pitch angle evolution.

Numerical simulations suggest that strong Alfv\'enic turbulence, particularly in magnetically dominated regimes — where the energy of magnetic fluctuations exceeds the rest mass energy of plasma particles — can {efficiently} accelerate particles to suprathermal energies \cite[e.g.,][]{zhdankin2017a,comisso2019,comisso2022,zhdankin2020,nattila2020,nattila2022,vega2022a,vega2023,vega2024,vega2025}. This phenomenon may be linked to the generation of shocks, reconnecting current sheets, magnetic mirrors, and other nonlinear structures, which are known to play a significant role in astrophysical particle acceleration \cite[e.g.,][]{matthaeus_turbulent_1986,blandford1987,uzdensky2011,drake2013,sironi2014,marcowith2016,loureiro2017,loureiro2018,loureiro2019,mallet2017a,boldyrev_2017,boldyrev_loureiro2018,boldyrev2019,walker2018,comisso2019,roytershteyn19,guo2020,guo2023,trotta2020,ergun2020a,lemoine2021,lemoine2023,pezzi2022,bresci2022,sironi2022,dong2022,french2023,vega2020,vega2022a,xu2023,das2025}.

Numerical studies of {particle} acceleration in a {turbulent} pair plasma {have demonstrated} that the {strength of the mean magnetic} field significantly affects the acceleration process. At weak guide fields, $B_0\ll \delta B_0$, the {power-law spectrum of non-thermal electrons appears to be slightly steeper} than $f(\gamma)d\gamma\propto \gamma^{-2}d\gamma$ {(e.g., \cite[][]{vega2022a} found the exponent to be $-2.1$ for $B_0 \sim \frac{1}{4} \delta B_0$). As the guide field $B_0$ increases, the lower bound of the energy spectrum's power-law tail appears at progressively larger energies~$\gamma$ and with progressively steeper exponents.} As was shown in \cite[][]{vega2022a}, already at $B_0\sim 3\,\delta B_0$ the power-law tail can hardly be discerned in the {steeply} declining energy distribution function. At even {greater} guide fields, the power-law tail disappears, and the electron energy distribution function is well approximated by a log-normal shape \cite[][]{vega2024}. 

This behavior is consistent with the conservation of the electron's magnetic moment 
prohibiting the increase of the field-perpendicular component of the electron momentum. The {field-parallel} acceleration depends on the relatively slow particle drifts and, as argued in \cite[][]{vega2025}, leads to a log-normal energy distribution of the accelerated electrons. The constraint imposed on the pitch-angle evolution by the conservation of the magnetic moment thus has a profound effect on the acceleration process.  

In this work, we {first} study the {energy-dependence of }pitch-angle distribution functions of accelerated electrons in PIC simulations of magnetically dominated turbulence in a pair plasma. We consider the case of a relatively strong guide field, $B_0/\delta B_0=10$, and large initial magnetization {(defined in Eq. \ref{sigma_tilde}),} $\tilde{\sigma}_0\sim 40$. 
{In Section 3}, starting from isotropic initial angular distributions of mildly relativistic electrons, we study the {dynamically evolved pitch angles of particles accelerated by decaying turbulent fluctuations}. We find that as the particles' gamma factors increase, their {average} pitch angles decrease {at a steep rate}. At higher energies (about $\gamma\sim 200$ in our case), the pitch-angle{s decline less rapidly as a function of $\gamma$}, consistent with {the phenomenological model in }~\cite[][]{vega2024b,vega2025}. The energy distribution function of accelerated particles declines rapidly with increasing Lorentz factor,~$\gamma$. As a result, the statistical ensemble of accelerated particles becomes quite limited at high energies. Specifically, it becomes challenging in our studies to analyze the angular distributions of particles for energies beyond $\gamma \sim 200$. 

To overcome these numerical challenges and extend our analysis to significantly higher energies, we adopt a second approach {in Section 4}: we examine the angular distributions of test particle {trajectories} in stationary magnetic and electric fields obtained from PIC simulations. {When the electric field is neglected, this} method allows us to analyze particle angular distributions up to {arbitrary energies due to our freedom to prescribe the initial Lorentz factor of test particles}. We find a reasonably good agreement with the phenomenological model presented in \cite[][]{vega2024b,vega2025}. {In particular, we find evidence for the four predicted stages of pitch angle evolution as a function of $\gamma$. The initial stage is characterized by pitch angle focusing that scales as $1/\gamma$. Upon reaching a minimum value set by the intrinsic curvature of the magnetic field, the pitch angle broadens with $\gamma$, until the particle's gyroradius is large enough to interact with turbulence, leading to broadening that scales as $\gamma^{1/2}$. Finally, for very high $\gamma$, the pitch angle reaches a saturated state with $\sin{\theta} \sim \delta B_0/B_0$. We emphasize that pitch angles are predicted to remain small throughout their evolution, never reaching the isotropic distribution often assumed for the interpretation of synchrotron spectra.}

Simultaneously, our test-particle studies reveal significant numerical limitations that are specific to {the} considered case of a strong guide field. In such situations, the particle's magnetic moment is well preserved, meaning that the particle's gyroradius does not change significantly during the acceleration process. In PIC simulations with a strong guide field, the particle's gyroradius may remain comparable to or even smaller than the numerical cell size for a considerable portion of the acceleration process. 

We have found that this scenario can negatively impact angular collimation due to pitch-angle scattering caused by the numerical noise {residing at the cell size, which arises from having finite particles per cell in the simulation.}
Even when the noise is weak, nonphysical scattering becomes significant if pitch angles become small. 
We, however, demonstrate that increasing the number of particles per cell and filtering out the electromagnetic fluctuations related to numerical noise can help bring the pitch-angle values closer to analytical predictions. 

Another observed limitation is related to the interpolation procedure used to numerically reconstruct the magnetic field inside a cell.  We find that even when the magnetic field Fourier modes associated with the numerical noise are filtered out, the interpolation method still significantly affects the pitch angle evolution. The smoother the interpolated magnetic field, the better the agreement with analytical predictions. We hypothesize that when the transitions of the {interpolated} fields between cells are not smooth enough, the analytical assumptions regarding the curvature of the magnetic field lines, which are necessary to preserve the first adiabatic invariant, may be violated, consequently affecting the evolution of the pitch angle. {For example, interpolation schemes without $C^1$ continuity will not keep the second derivatives at cell interfaces finite, potentially causing the curvature experienced by particles to be ill-behaved.}

Our studies of the angular distributions of accelerated particles and their scalings with the particle energy may be valuable for the interpretation of synchrotron spectra of relativistic astrophysical sources.   

\section{Numerical setup}
We will analyze the results of three particle-in-cell simulations of decaying magnetically dominated turbulence with the fully relativistic code VPIC~\cite[][]{bowers2008}. The simulations are performed in a ``2.5D" geometry, {wherein} the uniform guide magnetic field, $\bm{B}_0$, is applied in $z$-direction, while the magnetic and electric fluctuations are varied only in the $x$-$y$ plane. 
The system thus has continuous translational symmetry along the $z$-direction. 

The code evolves all three vector components of the fields, and all three components of the momenta of the particles propagating in these fields. Such a setup is expected to capture the essential nonlinear interactions of fields and particles{, retaining the shear modes necessary for Alfv\'enic turbulence. This setup} is known to produce field energy spectra and particle energy distributions similar to those obtained in fully 3D simulations, while {allowing computational resources to be allocated to} significantly {greater} numerical resolution and number of particles per cell \cite[see, e.g.,][]{zhdankin2017a, zhdankin2018c, comisso2018, comisso2019,nattila2022,vega2023}.

The simulations are conducted in a double periodic $L\times L$ square domain. The turbulent fluctuations are initialized by imposing randomly phased large-scale perturbations of the magnetic field,  
\begin{eqnarray}
\label{deltaB0}
\delta{\bm B}(\mathbf{x})=\sum_{\mathbf{k}}\delta B_\mathbf{k}\hat{\xi}_\mathbf{k}\cos(\mathbf{k}\cdot\mathbf{x}+\phi_\mathbf{k}),
\end{eqnarray}
where the unit polarization vectors are chosen to be normal to the background magnetic field, $\hat{\xi}_\mathbf{k}=\mathbf{k}\times {\bm B}_0/|\mathbf{k}\times{\bm B}_0|$, in order to initiate the shear-Alfv\'en type fluctuations. The two-dimensional wave vectors of the modes, $\mathbf{k}=\{2\pi n_x/L,{2}\pi n_y/L\}$, are chosen within the interval $n_x,n_y=1,...,8$. All the modes in Eq.~(\ref{deltaB0}) have the same amplitudes $\delta B_{\mathbf{k}}$, but random phases~$\phi_\mathbf{k}$. The initial root-mean-square value of the perturbations is given by $\delta B_0={\langle |\delta{\bm B}(\mathbf{x})|^2 \rangle^{1/2} }$, where the average is done over the simulation domain.  The {nominal} outer scale of turbulence can then be defined as $l=L/8$. The corresponding time scale, $l/c$, where $c$ is the speed of light, will be used to normalize time in the numerical results. 

All runs employ the domain size of $23552^2$ cells, which approximately corresponds to $1600^2$ electron inertial scales. Here, $d_e=c/\omega_{pe}$ denotes the nonrelativistic electron inertial scale, $\omega_{pe}=\sqrt{4\pi n_0e^2/m_e}$ the nonrelativistic electron plasma frequency{, and $n_0$ the mean density of each species}.   In all runs, the relative strength of the guide field is~$B_0/\delta B_0=10$.

One can define two plasma magnetization parameters based on the strengths of the guide field and magnetic fluctuations: 
\begin{eqnarray}
\sigma_0=\frac{B_0^2}{4\pi n_0 w_0 m_e c^2},    
\end{eqnarray}
and
\begin{eqnarray}
\tilde{\sigma}_0=\frac{(\delta B_0)^2}{4\pi n_0 w_0 m_e c^2}, 
\label{sigma_tilde}
\end{eqnarray}
Here, $w_0 m_e c^2$ is the initial enthalpy per particle. The initial distributions of both the electrons and the positrons are chosen to have the isotropic Maxwell-J\"uttner form with the mildly relativistic temperature $\Theta_0=k_BT_e/m_ec^2=0.1$.  For such a distribution, the specific enthalpy is given by $w_0=K_3(1/\Theta_0)/K_2(1/\Theta_0)\approx 1.27$, 
where $K_\nu $ is the modified Bessel function of the second kind. 

We also define the relativistic inertial scale in the electron-positron plasma as $d_{rel}=\sqrt{w_0 c^2/(2\omega_{pe}^2)}$. The initial magnetizations in our simulations {are} $\sigma_0=4000$, and $\tilde{\sigma}_0=40$. The runs, however, employ different numbers of particles per cell and (or) different numerical time steps. The parameters of the runs are summarized in Table~\ref{table}. The {nominal relativistic} electron gyroradius, $\rho_0=m_ec^2/|e|B_0$, is then $\sqrt{\sigma_0}\approx 60$ times smaller than the inertial scale~$d_e$, which means that it is about 4~times smaller than the cell size.

\begin{table}[t!]
\vskip5mm
\centering
\begin{tabular}{c c c c c c} 
\hline
{Run} & size $(d_e^2)$ & \# of cells & $\omega_{pe}{\delta} t$  & \# ppc \\
\hline
I & $1600^2$ & $23552^2$ & $6.0\times10^{-3}$ & 50 \\ 
II & $1600^2$ & $23552^2$ & $6.0\times10^{-3}$ & 200 \\ 
III & $1600^2$ & $23552^2$ & $3.0\times10^{-3}$ & 50 \\ 
\hline
\end{tabular}
 \caption{Parameters of the PIC runs. {Here, $d_e=c/\omega_{pe}$ is the nonrelativistic electron inertial scale, and $\delta t$ is the numerical time step.}}
\label{table}
\end{table}

Our previous numerical studies \cite[][]{vega2024} demonstrated that when the initial magnetic perturbations relax, they excite, in addition to Alfv\'en modes, high-frequency ordinary modes {that constitute a small fraction of the turbulent energy}. This happens because the magnetic field fluctuations in the ordinary mode are also polarized in the $x$-$y$ plane, so they may be generated by the initial magnetic perturbations {from Eq.} ~(\ref{deltaB0}). The electric field of ordinary modes is, however, polarized mostly in the $z$-direction.   In order to ensure that the initial perturbations mostly consist of Alfv\'en modes, we introduce an initial plasma current characteristic of the Alfv\'en modes, $J_z=(c/4\pi){\bm \nabla}\times {\delta \bm B}_{\perp,0}$. This approach, also adopted in \cite[][]{vega2024,vega2025} reduces the fraction of ordinary modes that are generated by initial perturbations. 

To incorporate this initial ``compensating" current, we add a velocity $U_{z}^s = J_z/(2 q_s n_0)$ to each particle of species $s$, where $q_s = \pm |e|$ (representing positrons and electrons), as follows: $v^s_z\to v^s_z+U^s_z$. Here, the particle velocity ${\bm v}^s$ is sampled from the isotropic Maxwell-Jüttner distribution. This velocity adjustment is made under the condition that the resulting velocity satisfies $|{\bm v}^s + {\bm U}^s| < 0.97 c$, to avoid an artificial generation of very energetic particles. For particles where such an adjustment would lead to $|{\bm v}^s + {\bm U}^s| \geq  0.97 c$, we keep the particle velocity unchanged. As verified in \cite[][]{vega2025}, the addition of this current does not alter the core of the particle energy distribution function but adds a weak tail extending up to~$\gamma\approx 4$, which is not essential for our study of very energetic accelerated particles with $\gamma>20$. 

\section{Results of PIC simulations}
\begin{figure}[h!]
\centering
\includegraphics[width=1.05\columnwidth]{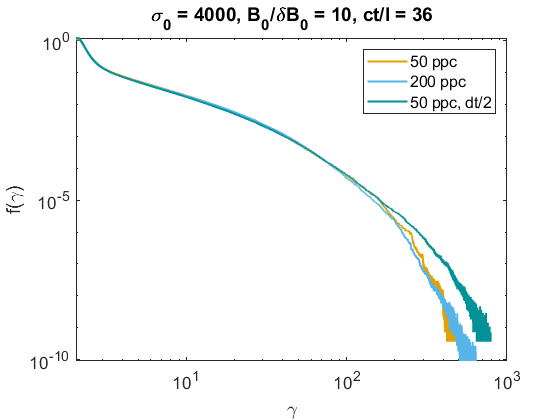}
\caption{Electron energy distribution functions in Runs~I, II,~and~III. 
\label{energy}}
\end{figure}

Figure~\ref{energy} illustrates the particle energy distribution functions in our simulations, measured at~$ ct/l = 36$, which corresponds to approximately $3.6 $~large-scale eddy crossing times (it is important to note that the field-perpendicular velocity associated with large-scale Alfvénic fluctuations is roughly $0.1c $). By around $ 3 $ to $ 4 $ {eddy} crossing times, nearly half of the energy contained in the initial magnetic perturbations has been transferred to the particles. Consequently, the fluctuation magnetization has decreased to $ {\tilde \sigma}_0 \approx 1 $. The particle energy distribution functions have reached a quasi-saturated state, and the average {Lorentz factor} of the particles has risen to $ \langle \gamma \rangle \approx 6 $. Continuing the simulation for longer periods does not significantly alter the particle distribution function, as the energy {available} from the initially strong magnetic fluctuations has diminished substantially, preventing further efficient energization of the particles.
As discussed in \cite[e.g.,][]{vega2025}, the energy distribution function can be accurately approximated by a log-normal shape at suprathermal energies,  $\gamma \gg 6$. 

It is important to note that our three simulation runs are physically identical; they differ only in terms of numerical accuracy, which is determined by the number of particles per cell and the {temporal resolution}. We observe that the range where the curves in Figure~\ref{energy} overlap extends up to energies of approximately $\gamma \sim 200$; this range is where our numerical results can be considered reliable. At higher energies, discrepancies between the curves appear. These discrepancies are, therefore, due to numerical effects as well as limitations in the statistical ensemble, which results from the relatively small number of particles accelerated to such high energies. 

\begin{figure}[h!]
\centering
\includegraphics[width=1.1\columnwidth]{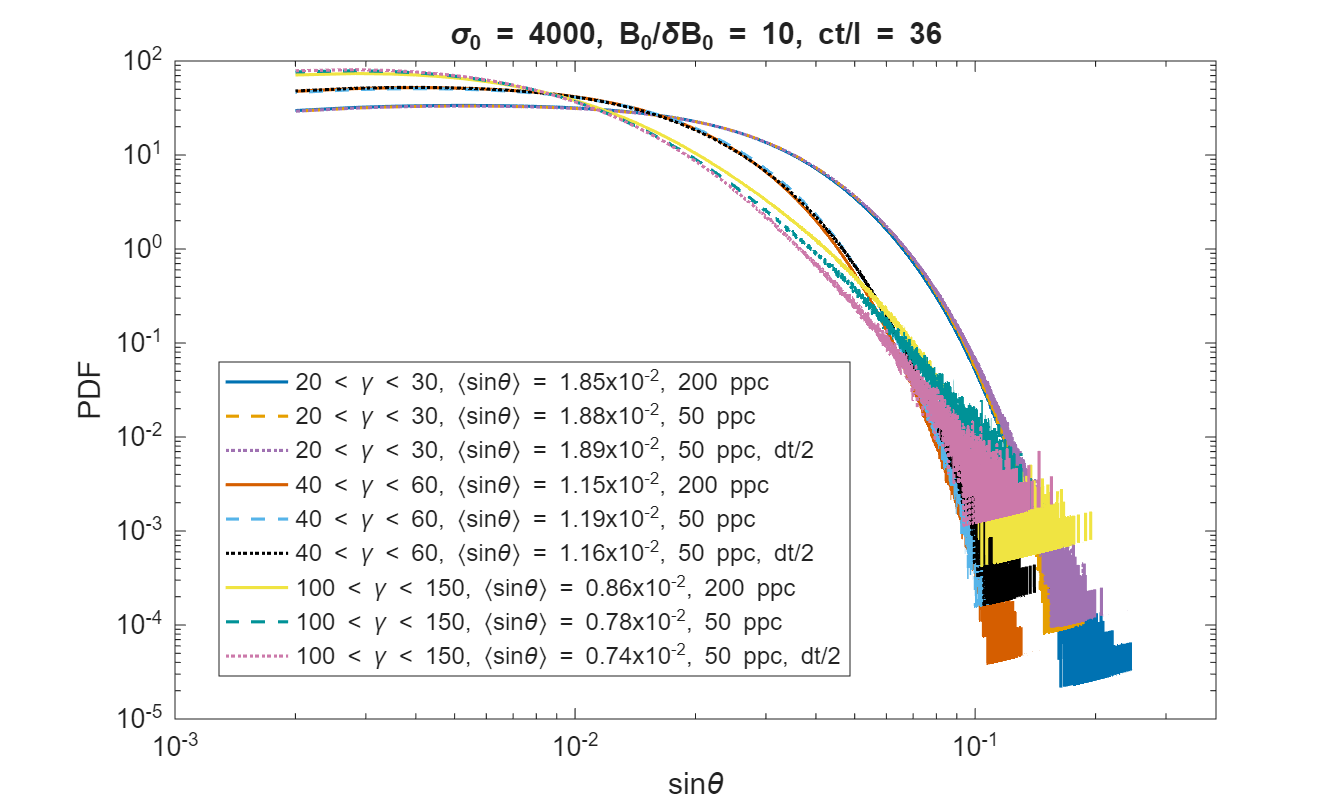}
\caption{Pitch angle distributions corresponding to different energy intervals of accelerated particles, measured in Runs~I,~II, and~III. The pitch angle is calculated in the local plasma frame, which is co-moving with the local~``$E\times B$'' velocity.
\label{sin_theta}}
\end{figure}

\begin{figure}[h!]
\centering
\includegraphics[width=1.1\columnwidth]{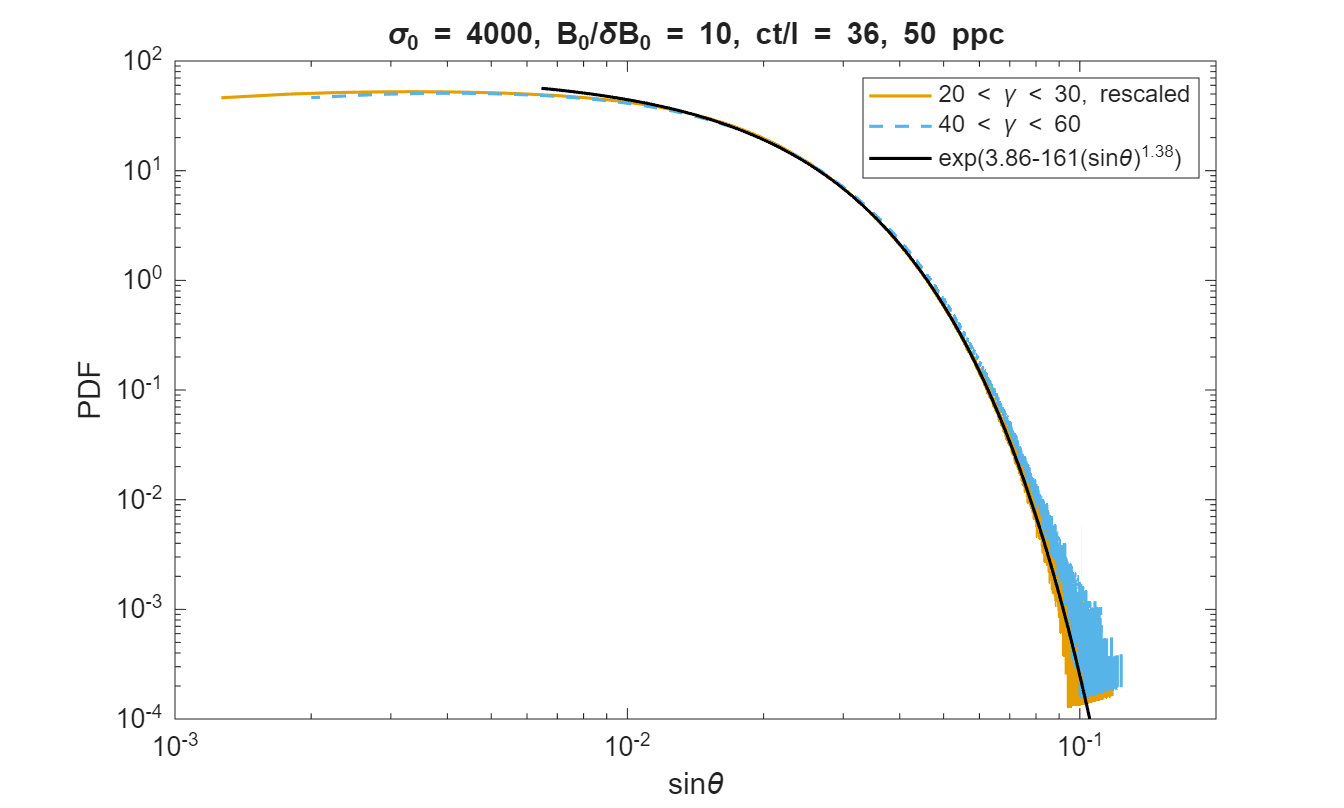}
\caption{Compressed-exponential fit to the tail of the self-similar angular distribution function{, with empirically fitted parameters $\delta\approx 1.38$, $\lambda \approx 161$, and $A \approx 3.86$ (Eq. \ref{compexp_eq})}. The argument in the angular distribution function corresponding to $20<\gamma <30$ is rescaled according to $\sin\theta\to (\sin\theta)/1.6$, to overlap with the case of the larger $\gamma$. 
\label{sin_theta_fit}}
\end{figure}

We will now analyze the angular distributions of particles accelerated to different energies. As shown in Figure~\ref{sin_theta}, the angular distribution functions obtained in different runs for energies below $\gamma \sim 80$ overlap and evolve in a similar fashion. As {was found} in \cite[][]{vega2025}, these distribution functions display a self-similar form.  When the $\gamma$-factor doubles, the angular distribution function narrows by approximately a factor of~$1.6$. Figure~\ref{sin_theta_fit} shows this self-similar behavior and demonstrates that the {tail of the} measured self-similar {angular distribution for a given value of $\gamma$} can be {approximated} by a compressed-exponential function,
\begin{eqnarray}
f(\sin\theta)\,d\sin\theta =e^{{A}-{\lambda}(\sin\theta)^\delta }\,d\sin\theta,  
\label{compexp_eq}
\end{eqnarray}
with {empirically fitted parameters $\delta\approx 1.38$, $\lambda \approx 161$, and $A \approx 3.86$.}

However, at higher energies where $\gamma > 100$, we start observing deviations from this self-similar behavior; the {tails of the angular distributions for $100 < \gamma < 150$ in Fig. \ref{sin_theta} becomes overly broad}. This change may indicate that as the pitch angles become smaller, the angular distribution is influenced by different mechanisms, such as the slow drifts caused by the curvature and gradient of the local magnetic field \cite[][]{vega2025}. Studying this modified behavior is challenging in our simulations since at such energies, the curves from different runs start to deviate from each other. These discrepancies are likely due to numerical effects and/or an insufficient statistical ensemble, as previously discussed.

In an attempt to overcome these numerical challenges and get some insight about the angular distributions at larger energies, we will analyze the evolution of test particles in {\em stationary} electric and magnetic fields obtained from {our PIC simulations at the stage of fully-developed turbulence (around $ct/l = 36$)}. Test particle methods have been used in many previous studies of particle acceleration within turbulent magnetic and electric fields, in both dynamic and stationary settings \cite[e.g.,][]{miller1997,giacalone1999,ergun2020a,ergun2020b, lemoine2023b, kempski2023}.

\section{Test particle simulations}
We will now study the angular distributions of test particles accelerated to different energies in electric and magnetic fields obtained from PIC simulations.  We utilize a standard relativistic Boris push \cite[][]{BirdsallLangdon1991} and bilinear field interpolation to integrate particle trajectories in stationary magnetic and electric fields. This approach would, of course, provide a simplified model for particle acceleration, as a realistic electromagnetic field is not stationary. Instead, turbulent fluctuations renovate on the order of a large-scale eddy turnover time, which is approximately $c\Delta t/l \sim 10$ in our case. However, we expect that by tracing test particles for only a few large-eddy crossing times, we can effectively investigate particle acceleration within the eddies. Tracing for longer periods may become unphysical, as the results could be distorted by particle trapping and acceleration in persistent strong structures.

\begin{figure}[h!]
\centering
\includegraphics[width=1.1\columnwidth]{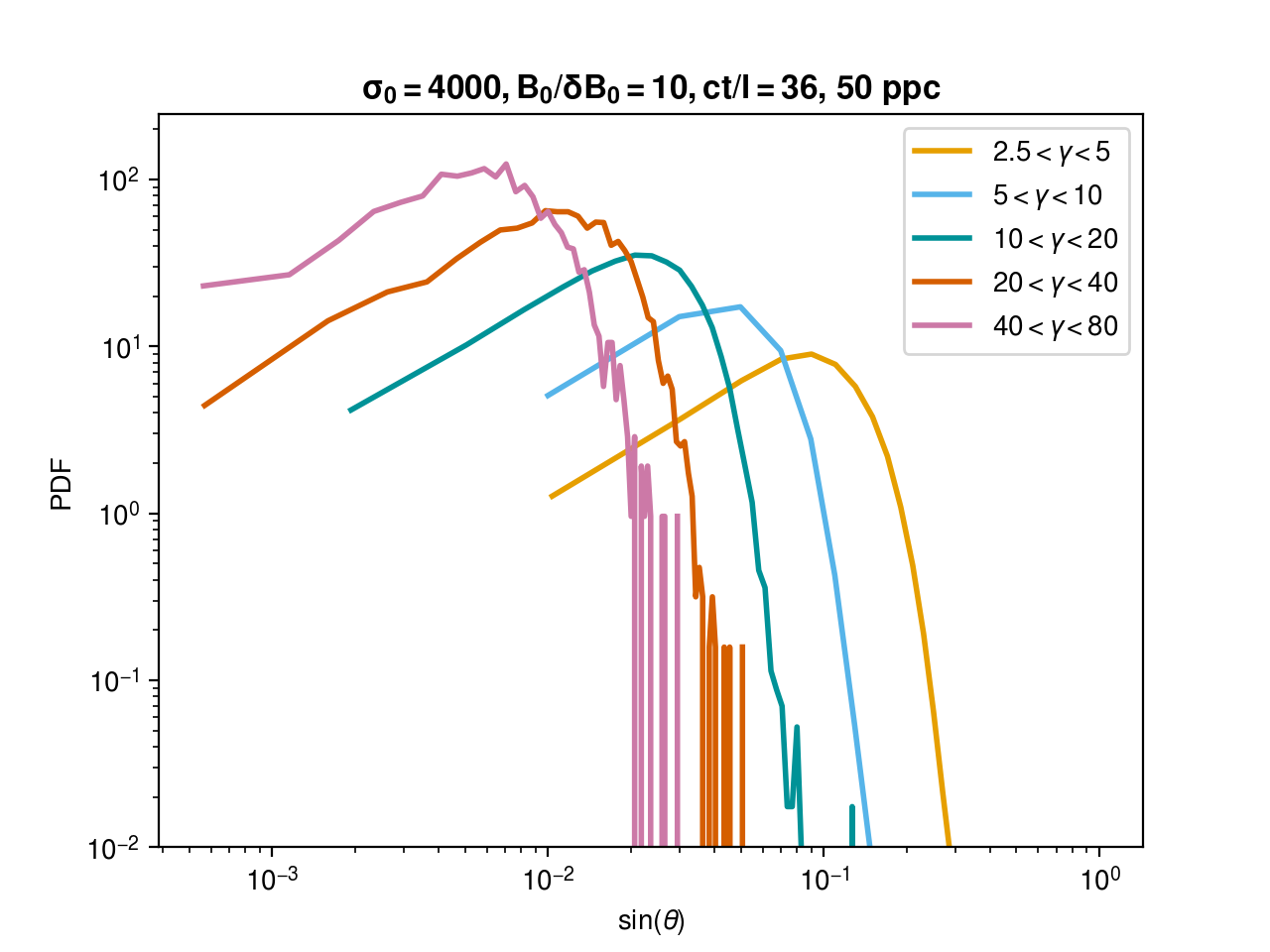}
\caption{{Pitch angle distribution for particles accelerated in static fields. Particles were traced over the time interval of~$c\Delta t/l=2$. The pitch angles are calculated in local plasma frames, co-moving with the local~``$E\times B$'' velocities. The distributions for different $\gamma$ exhibit a self-similar nature.}
\label{pitch_dist_2ctl_unscaled}}
\end{figure}

\begin{figure}[h!]
\centering
\includegraphics[width=1.1\columnwidth]{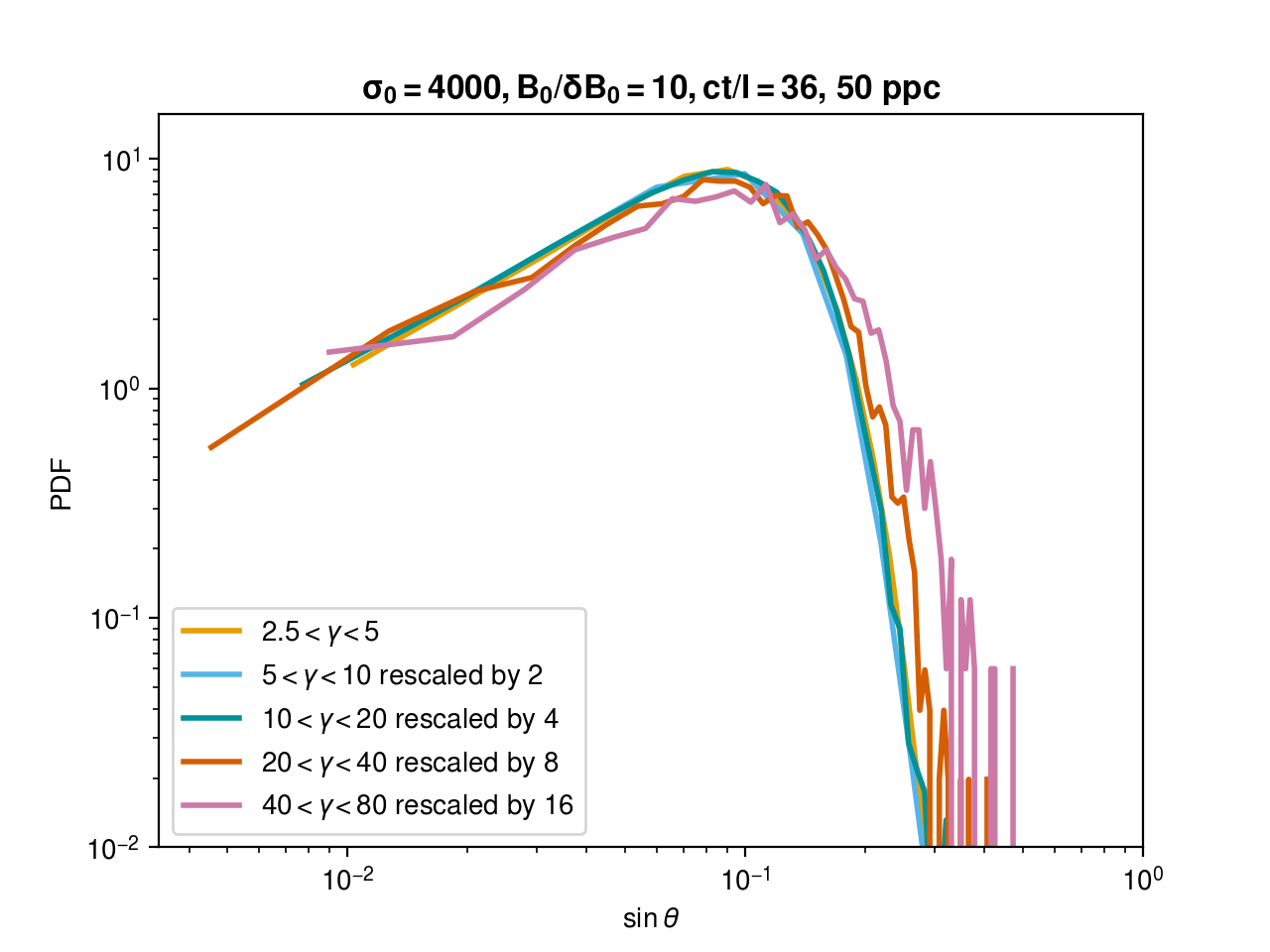}
\caption{Pitch angle distribution for particles accelerated in static fields. The curves are rescaled to overlap with the case of the lowest $\gamma$ value, {according to the theoretically motivated $\sin \theta\sim 1/\gamma$ rescaling symmetry}. Particles were traced over the time interval of~$c\Delta t/l=2$. The pitch angles are calculated in local plasma frames, co-moving with the local~``$E\times B$'' velocities. Increasing deviations from theory {(i.e., imperfect overlap)} are present for higher $\gamma$ values.
\label{pitch_dist_2ctl}}
\end{figure}

We start with a {distribution of electrons with isotropic pitch angles and identical energies corresponding to the semi-relativistic velocity $v/c = 0.6$}, {which we trace over the} time interval~$c\Delta t/l{=} 2$ 
in the electric and magnetic fields taken from Run~I.  As seen in Figure~\ref{pitch_dist_2ctl_unscaled}, the {electron pitch-angle distributions exhibit self-similarity}, which broadly aligns with the results from the full PIC simulations. 
{In Figure \ref{pitch_dist_2ctl}, we show that} the {$\gamma$-dependent rescaling symmetry of the test particle pitch-angle distribution function} agrees more closely (although still not perfectly) with the analytical prediction {$\sin \theta\sim 1/\gamma$ \cite[][]{vega2025}} than the scaling obtained from the full PIC simulations. 

\begin{figure}[h!]
\centering
\includegraphics[width=1.1\columnwidth]{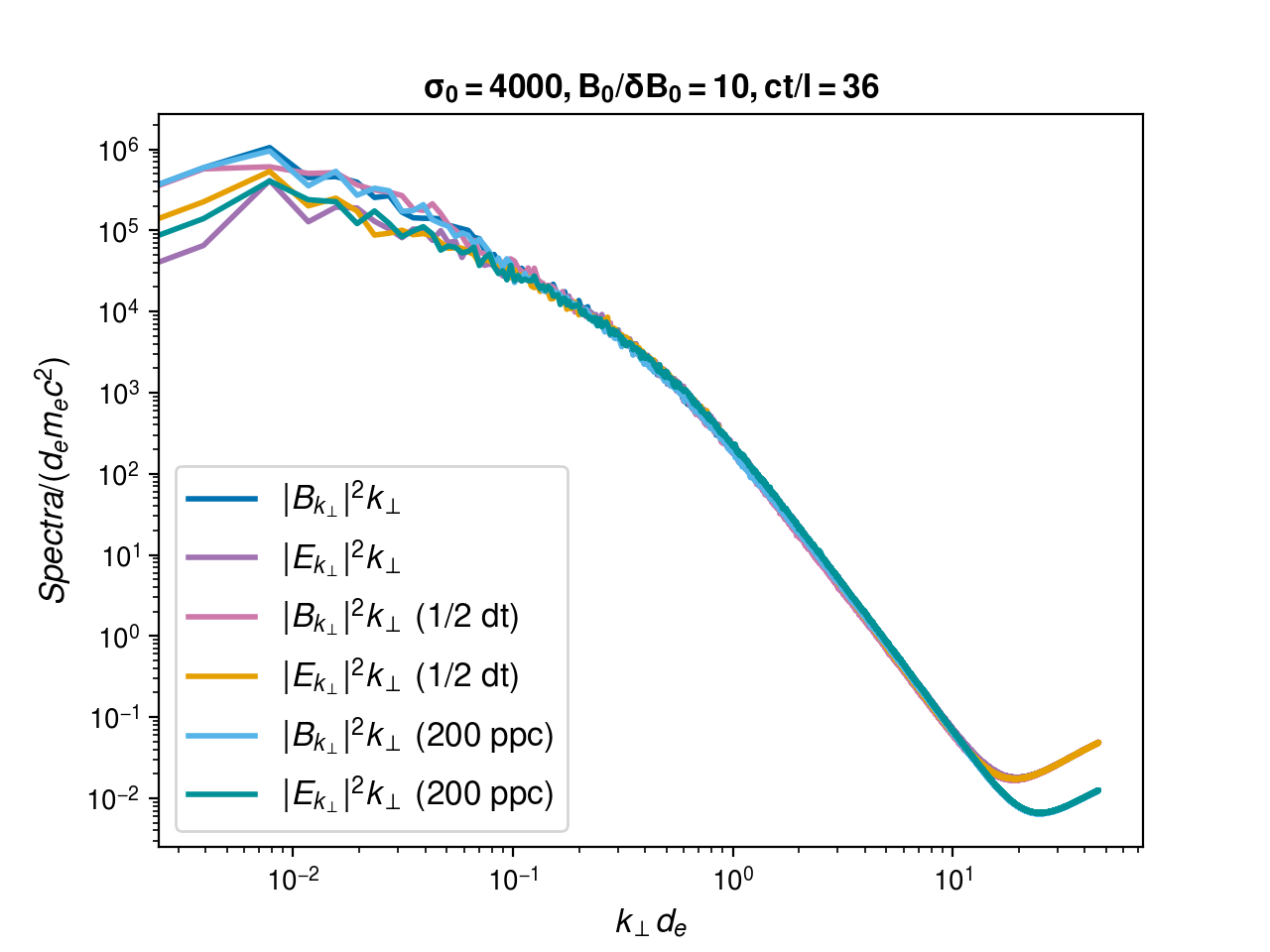}
\caption{Perpendicular magnetic and electric spectra for Run I, Run II, and Run III. The noise contributions appear as the rising parts in the Fourier energy spectra of the electric and magnetic fields at large wave numbers. To ``smooth" the fields, we removed harmonics associated with the noise, that is, harmonics greater than the minima of the spectra, about $kd_e = 20$.  
\label{spectra}}
\end{figure}

\begin{figure}[h!]
\centering
\includegraphics[width=1.1\columnwidth]{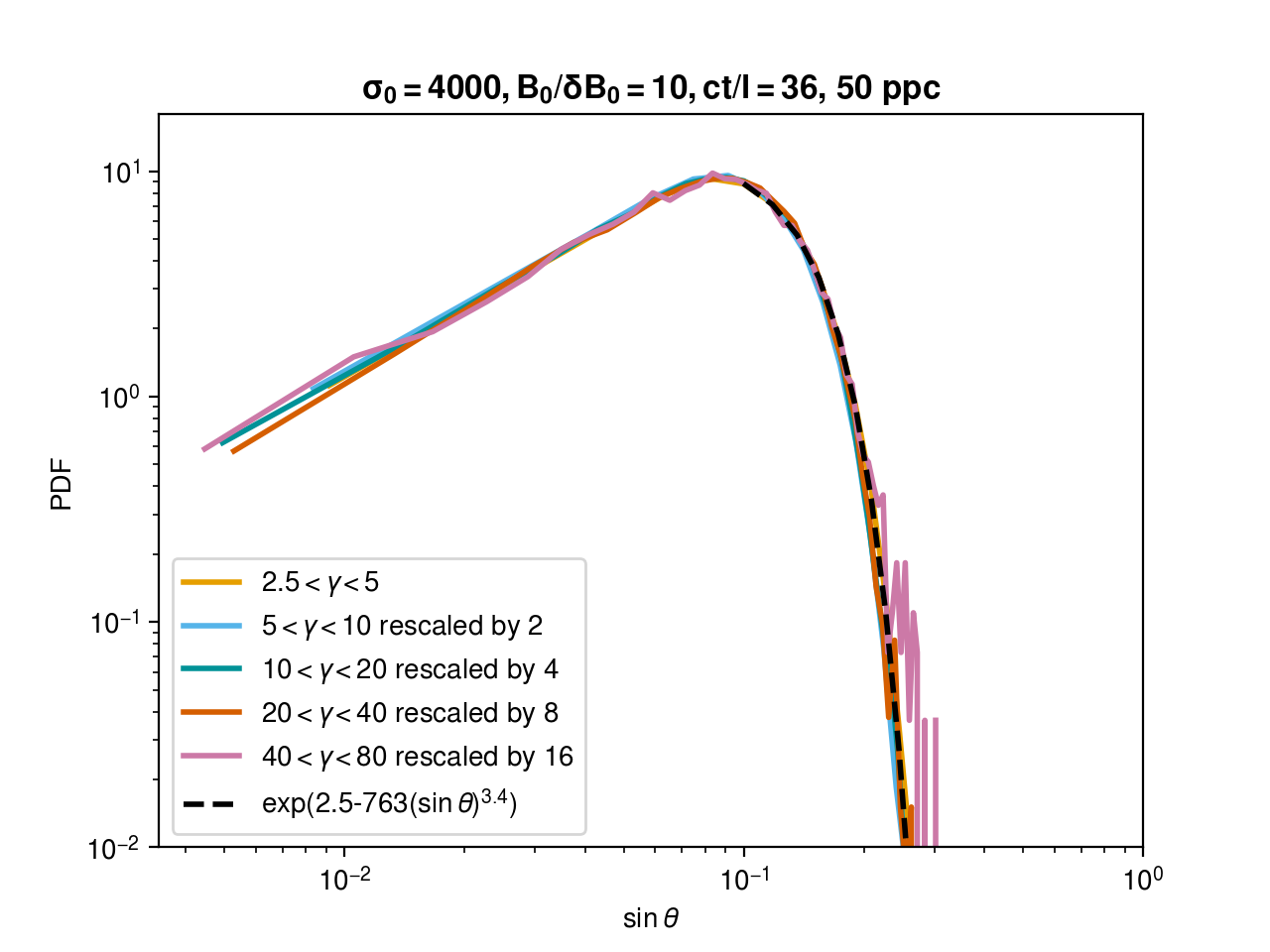}
\caption{Pitch angle distribution for particles accelerated in static, smoothed electric and magnetic fields. The curves are rescaled to overlap with the case of the lowest $\gamma$ value, {according to the theoretically motivated $\sin \theta\sim 1/\gamma$ rescaling symmetry}. Particles were traced over the time interval of~$c\Delta t/l=2$. The pitch angles are calculated in local plasma frames, co-moving with the local~``$E\times B$'' velocities. Agreement with theory is significantly improved, and the tail can be {approximated} by a compressed-exponential with {empirically fitted parameters} $\delta \approx 3.4${, $\lambda \approx 770$, {and $A\approx2.5$ (Eq. \ref{compexp_eq})}}. 
\label{pitch_dist_2ctl_smoothed}}
\end{figure}

We then repeat this test particle tracing procedure by using ``smoothed'' magnetic and electric fields, where the effects of PIC numerical noise have been reduced by removing the high-$k$ Fourier harmonics corresponding to the noise interval (Fig. \ref{spectra}). 
As seen in Figure~\ref{pitch_dist_2ctl_smoothed}, the self-similarity is now restored, and the scaling agrees well with the analytical prediction and extends to higher energies. The compressed exponential still provides a good fit to the tails of the distribution functions, albeit with a larger exponent~$\delta$ than in Fig.~\ref{sin_theta_fit}.

\begin{figure}[h!]
\centering
\includegraphics[width=1.1\columnwidth]{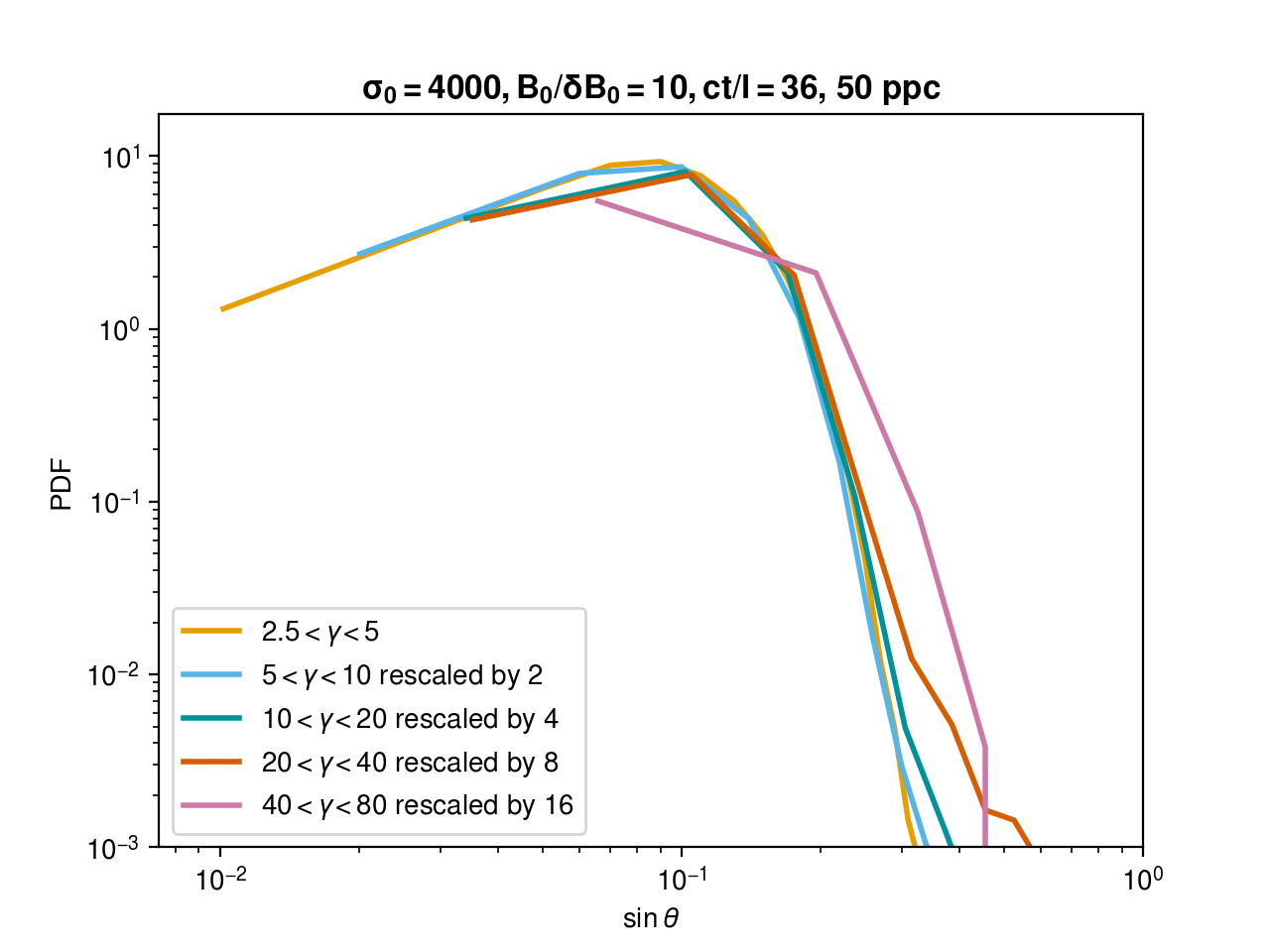}
\caption{Pitch angle distribution for particles accelerated in static, smoothed electric and magnetic fields. {The curves are rescaled to overlap with the case of the lowest $\gamma$ value, according to the theoretically motivated $\sin \theta\sim 1/\gamma$ rescaling symmetry}. Particles were traced over the time interval of~$c\Delta t/l=5$. The pitch angles are calculated in local plasma frames, co-moving with the local~``$E\times B$'' velocities. Due to noise, the curves begin deviating from the analytical scaling, even in the smoothed case.
\label{pitch_dist_5ctl_smoothed}}
\end{figure}

Therefore, we suggest that the numerical noise {present} in PIC simulations, though weak, may have a significant effect on pitch angle scattering, particularly in the limit of small pitch angles. This suggestion is further supported by our observation that when we increase the tracing time to $c\Delta t/l = 5$ (Figure \ref{pitch_dist_5ctl_smoothed}), we once again begin to see deviations from the analytically predicted scaling, even in the ``smoothed" fields. This indicates that some pitch angle scattering caused by the residual noise remains active, and it becomes noticeable at longer integration times.

In order to study the pitch angle scattering provided by the small-scale PIC noise in more detail, we conducted the following experiment. We trace the particle evolution taking into account only the stationary magnetic field, that is, {\it fully neglecting the electric force}. In this case, the particle's energy is conserved. We conducted a series of test-particle runs for various energies of the test particles. In each run, we chose the particles' initial positions randomly. However, we chose their initial velocities to {\it align exactly} with the directions of their local magnetic field. In this case, all particles begin with a pitch angle of zero. 

Consequently, we expect that the variations in pitch angle would result exclusively from inherent magnetic curvature and gradient drifts. This calculation determines the best possible alignment that can be achieved between particle momentum and the magnetic field, regardless of any acceleration mechanisms. A particle with a specific energy cannot achieve alignment with the magnetic field with less uncertainty.

The analytical consideration presented in \cite[][]{vega2024b,vega2025} predicts that {at a critical $\gamma_c$, a particle's gyroradius will become comparable to the electron inertial scale $d_{rel}$, allowing its magnetic moment to be changed by interactions with turbulent eddies. Before this regime is reached,} $\gamma\ll\gamma_c$, the angular broadening scales as
\begin{eqnarray}
\label{rho_perp_small}
\sin\theta \sim {2}\gamma\left(\frac{\rho_0}{l} \right)\left(\frac{\delta B_0}{B_0}\right)^2 \left(\frac{l}{d_{rel}} \right)^{1/3},
\end{eqnarray}
where $\rho_0=m_ec^2/|e|B$, and 
\begin{eqnarray}
\label{gamma_c_orig}
\gamma_c\sim {\frac{1}{\sqrt{2}}}\frac{B_0}{\delta B_0}\frac{l}{\rho_0}\left(\frac{d_{rel}}{l}\right)^{2/3}.    
\end{eqnarray}
For the parameters of our simulations, we have order-of-magnitude estimates:
\begin{eqnarray}
\label{theta_estimate}
\sin\theta\sim {8.8\times 10^{-6}}\gamma,   
\end{eqnarray}
and 
\begin{eqnarray}
\label{gamma_c}
\gamma_c\sim {2.5}\times 10^3.    
\end{eqnarray}
{{For these estimates, we use the following values: $\delta B_0/B_0=0.1$, $l/d_{rel}\approx 250$, and $d_{rel}/\rho_0=\sqrt{w_0^2\sigma_0/2}\approx 56.8$. We, however, note that formulas~(\ref{theta_estimate}) and~(\ref{gamma_c})  should be viewed as approximate relations, as they are based on phenomenological estimates for magnetic fluctuations, eddy sizes, and eddy anisotropy that are not rigorously derived.}} 
{We also note that $\gamma_c$ is much greater than the average Lorentz factor of the fully developed turbulence, $\langle \gamma \rangle \approx 6.$}

\begin{figure}[h!]
\centering
\includegraphics[width=1.05\columnwidth]{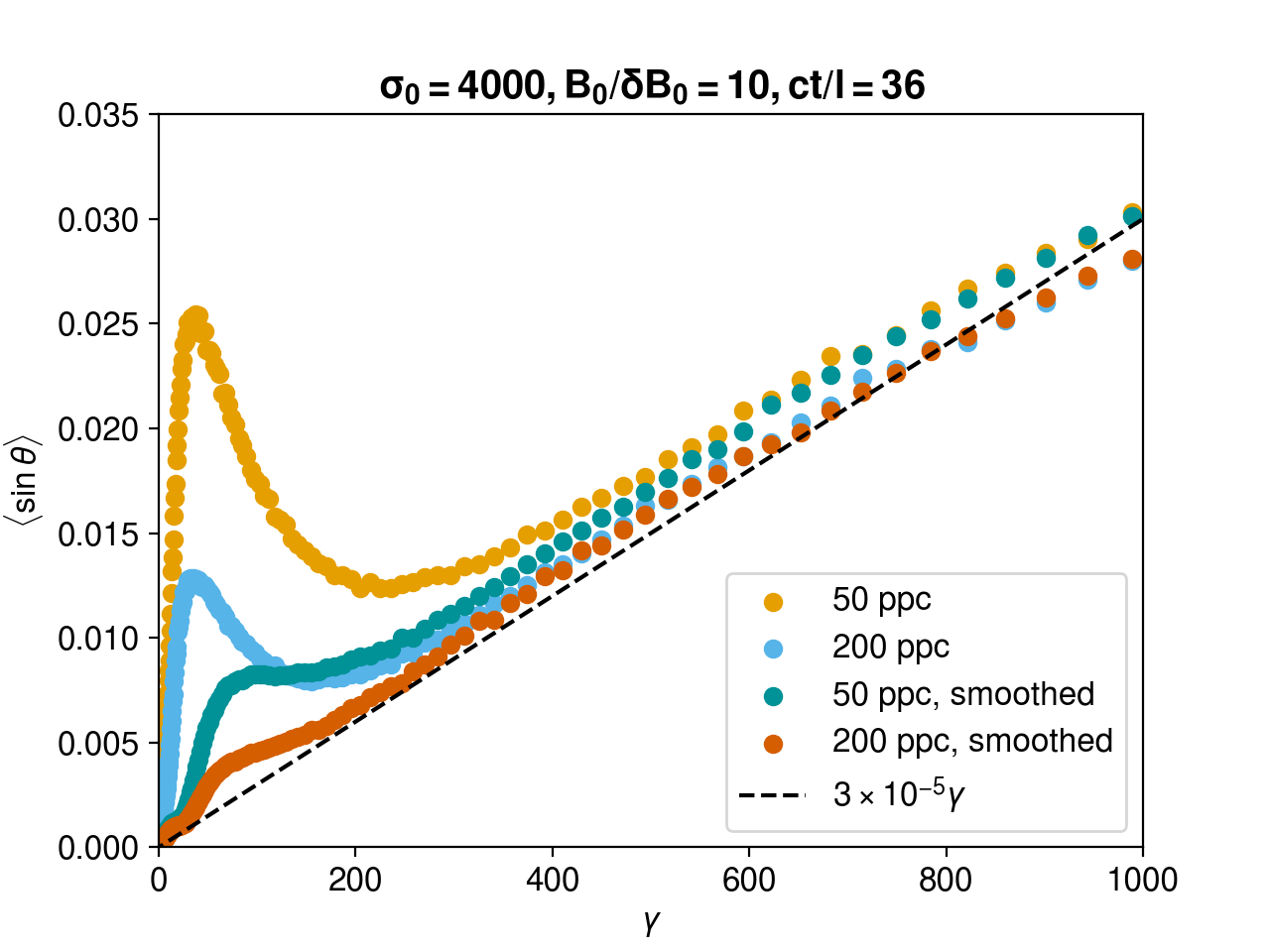}
\caption{Pitch angles of tracer particles measured after integrating particle trajectories in static magnetic fields for $c\Delta t/l=40$.  Each dot represents the average pitch angle of $5,000$~tracer particles corresponding to a specific energy,~$\gamma$. The pitch angle as a function of $\gamma$ approaches a linear fit as noise is reduced, as expected. The pitch-angle broadening is crucially affected by the numerical noise at $\gamma\lesssim 300$.
\label{noise_comparison}}
\end{figure}

The results of our test particle simulations are shown in Figure~\ref{noise_comparison}. We conducted simulations using the complete magnetic fields obtained in Runs~I and~II, as well as their ``smoothed" versions, where the high-frequency components of the spectra—associated with the numerical noise—were removed as discussed in Figure~\ref{spectra}. For each simulation, we integrate the trajectories of test particles over time~$c\Delta t/l=40$, allowing the particles to trace approximately four large-scale turbulent eddies. We logarithmically space particles' initial~$\gamma$ values from 1 to 10,000 in order to best resolve the effects of noise at low~$\gamma$. We examine $200$ distinct values of $\gamma$, probing each energy value with 5,000 particles.

We see from Figure~\ref{noise_comparison} that in all cases, the angular broadening at energies $\gamma \lesssim 300$ is primarily influenced by pitch-angle scattering due to unphysical numerical noise. However, at larger $\gamma$ the scaling is consistent with the theoretical prediction. The dashed straight line has a slope close to $\sin\theta\sim 3\times 10^{-5}\gamma$, which is not far from the analytical order-of-magnitude estimate given by~Eq.~(\ref{theta_estimate}).

\begin{figure}[h!]
\centering
\includegraphics[width=1.05\columnwidth]{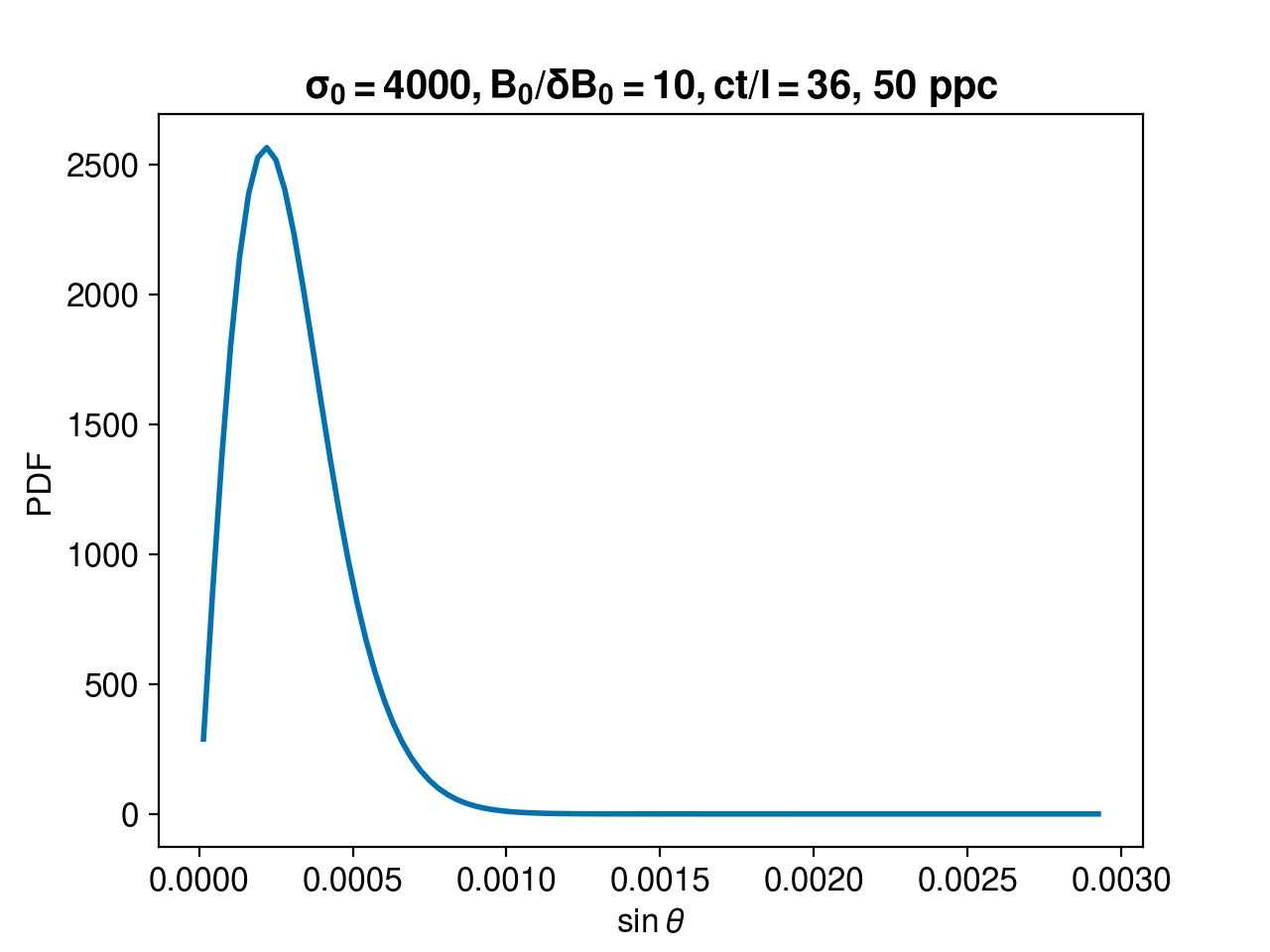}
\caption{ {Distribution of local angles between the original magnetic field $\mathbf{B}$ and the smoothed magnetic field $\mathbf{B}_s$ (i.e., $\sin{\theta} = {\mathbf{B}\times\mathbf{B}_{s}}/{\left|\mathbf{B}\right|\left|\mathbf{B}_{s}\right|}$). The smoothed field was obtained by removing all harmonics larger than the Fourier spectrum minimum, at about $kd_e = 20$ (Fig. \ref{spectra}). The measured angular variations of the magnetic field lines are, therefore, due to noise.  They are much smaller than the change in average pitch angle of test particles caused by noise (Fig. \ref{noise_comparison}).}
\label{field_noise_comparison}}
\end{figure}

To verify that we are observing pitch-angle scattering instead of merely magnetic angle fluctuations caused by noise, we present the distribution of magnetic angle fluctuations associated with the noise in Figure~\ref{field_noise_comparison}. This figure illustrates the local angles between the complete and smooth versions of the magnetic field. We see that the magnetic-field angular fluctuations resulting from the noise are relatively small, on the order of $3\times 10^{-4}$. In contrast, the particle pitch angles measured in Figure~\ref{noise_comparison} at $\gamma\lesssim 300$ are orders of magnitude larger. This indicates that although numerical noise itself is minimal, it provides significant pitch angle scattering and impacts particle dynamics at small angles.

\begin{figure}[h!]
\centering
\includegraphics[width=1.1\columnwidth]{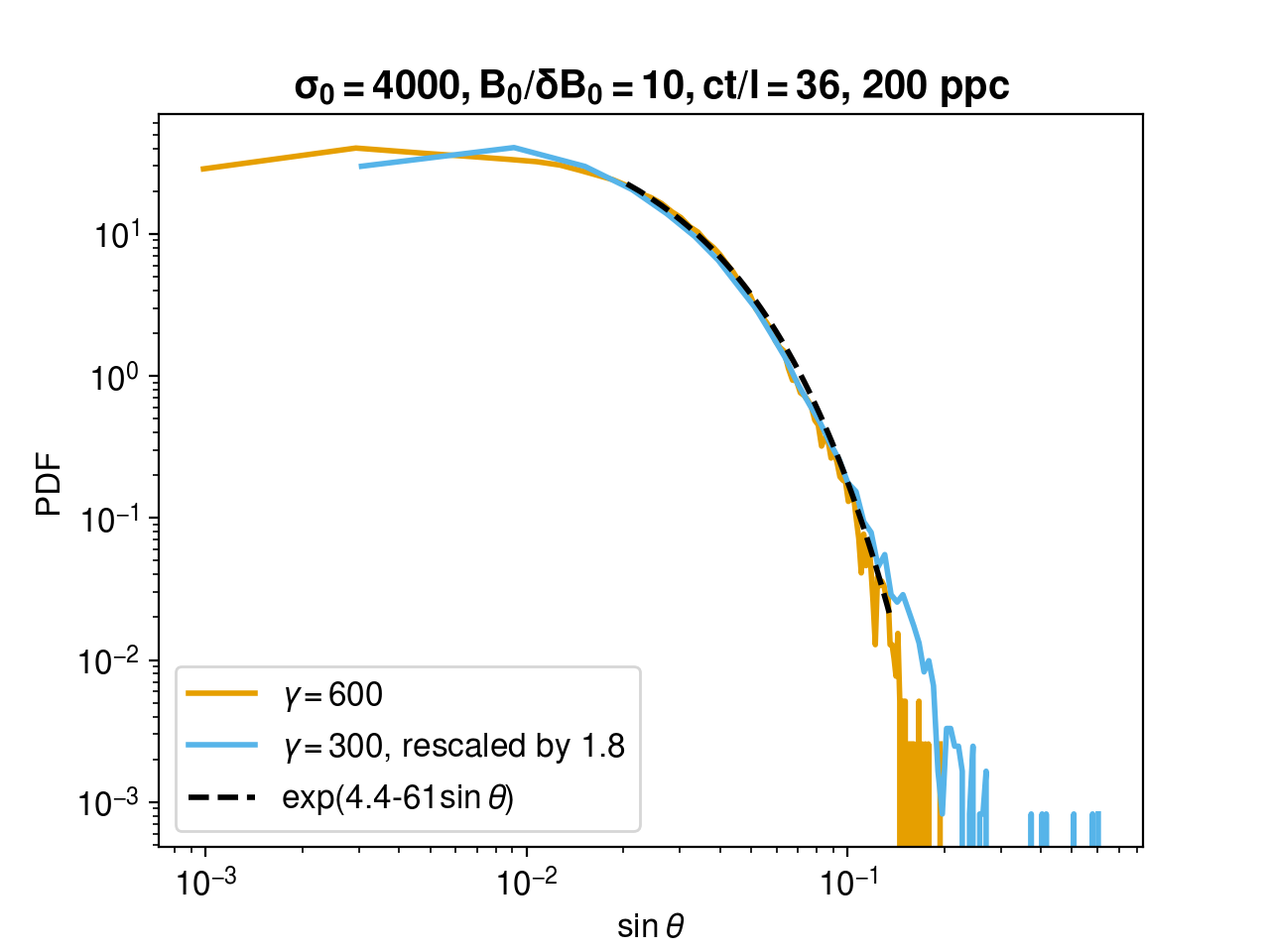}
\caption{  Distribution of pitch angles of tracer particles measured after integrating particle trajectories in static magnetic fields for $c\Delta t/l=40$, with $\gamma = 300$ and $600$. In order to better resolve the distribution functions, we used $2\times 10^5$ tracer particles for each value of~$\gamma$. The curve corresponding to $\gamma=300$ is rescaled by 1.8, showing close to linear rescaling symmetry (which would imply the rescaling factor of~$2$), see Eq.~\ref{theta_estimate}. The distribution is fit to an exponential curve. {The empirically fitted parameters are equivalent to a compressed exponential with $\lambda \approx 61$ and $A \approx 4.4$, but $\delta= 1$ (Eq. \ref{compexp_eq}).}
\label{g100_dist}}
\end{figure}

{We now study pitch-angle distributions for particles with energies $\gamma\gtrsim 300$, where the noise contamination is minimal.} As discussed in \cite[][]{vega2024b,vega2025} and illustrated in Figures~\ref{sin_theta} and~\ref{noise_comparison}, at these energies, the pitch angles are expected to be defined by the weak curvature drifts in the local magnetic field. Consequently, we attempt to examine them using our model of particle tracing in a stationary magnetic field provided by Run~II. The results are presented in Fig.~\ref{g100_dist}. We find that the angular distribution functions can be well fitted by an ordinary exponential shape, and their scaling with energy agrees broadly with the linear law given by the analytical prediction in Eq.~(\ref{theta_estimate}).

\begin{figure}[h!]
\centering
\includegraphics[width=1.1\columnwidth]{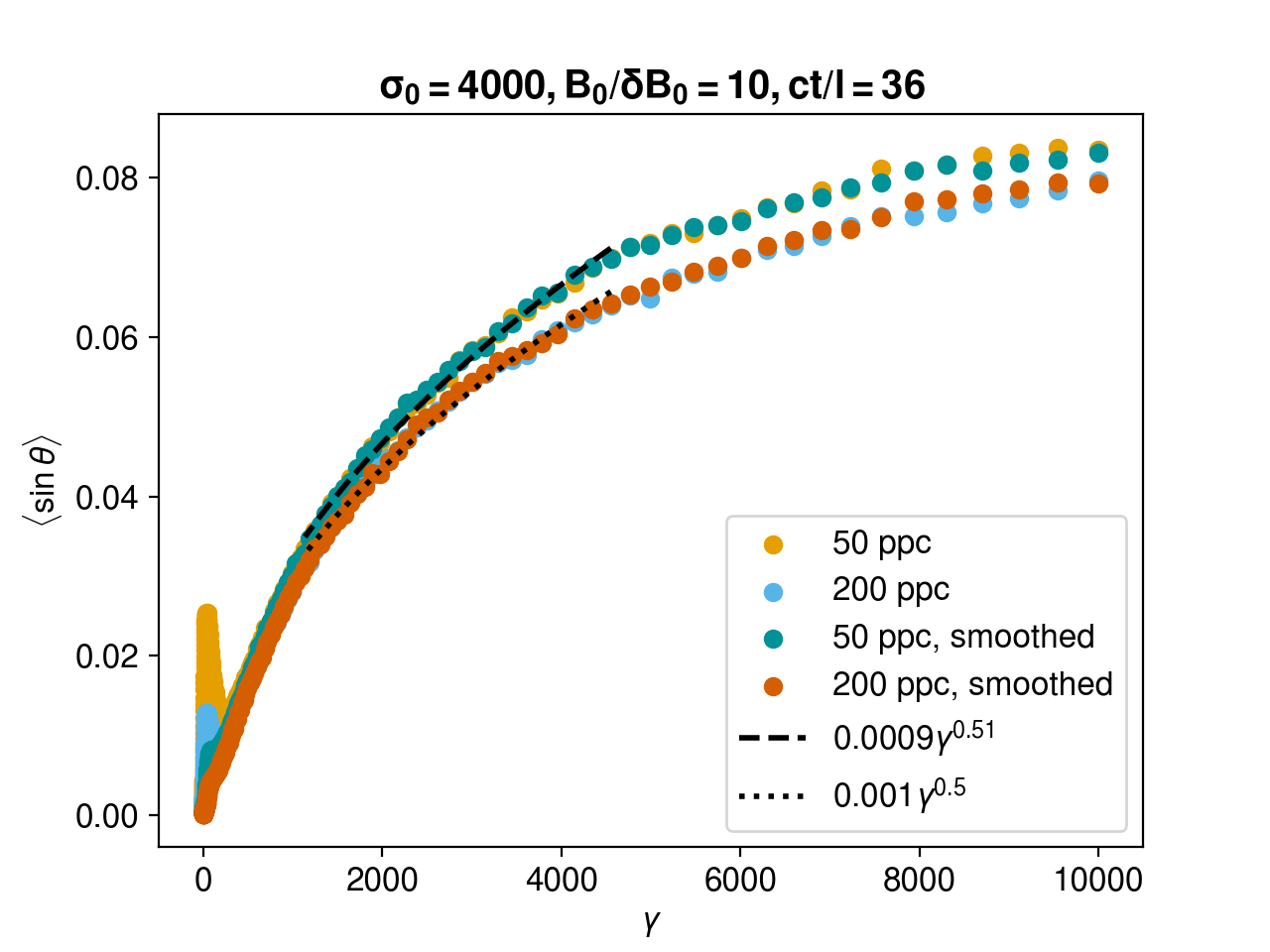}
\caption{Pitch angles of tracer particles measured after integrating particle trajectories in static magnetic fields for $c\Delta t/l=40$, with values of $\gamma$ stretching out to $\gamma = 10,000$. Each dot corresponds to a pitch angle value averaged over $5,000$ tracer particles. {We fit the curve to $\langle \sin{\theta}\rangle = C\gamma^b$, and find that in the intermediate range where $b$ is very close to $1/2$, the coefficient $C$ for both 50 and 200 ppc runs is only a factor of $\sim 2$ from Eq. \ref{sqrt_est}}. At large $\gamma$ the pitch angles begin to saturate and become independent of~$\gamma$.
\label{sqrt_fit}}
\end{figure}

At higher energies, $\gamma > \gamma_c$, where $\gamma_c$ is given by Eq.~(\ref{gamma_c_orig}), the analytical model \cite[][]{vega2025} predicts a square-root scaling for the pitch-angle broadening:
\begin{eqnarray}
\label{square_gamma}
\sin\theta \sim{2^{3/4}}\gamma^{1/2}\left(\frac{\delta B_0}{B_0} \right)^{3/2}\left(\frac{\rho_0}{l} \right)^{1/2}.    
\end{eqnarray}
For our parameters, we have an order-of-magnitude estimate:
\begin{eqnarray}
\label{sqrt_est}
\sin\theta\sim {4.5\times 10^{-4}}\gamma^{1/2}.    
\end{eqnarray}
Our tracer particle measurements shown in Figure~\ref{sqrt_fit} are broadly consistent with this estimate. The scaling behavior given by Eq.~(\ref{square_gamma}), however, begins at approximately $\gamma \sim 800-1000$, which are somewhat smaller values than those given in Eq.~(\ref{gamma_c}).

\begin{figure}[h!]
\centering
\includegraphics[width=1.1\columnwidth]{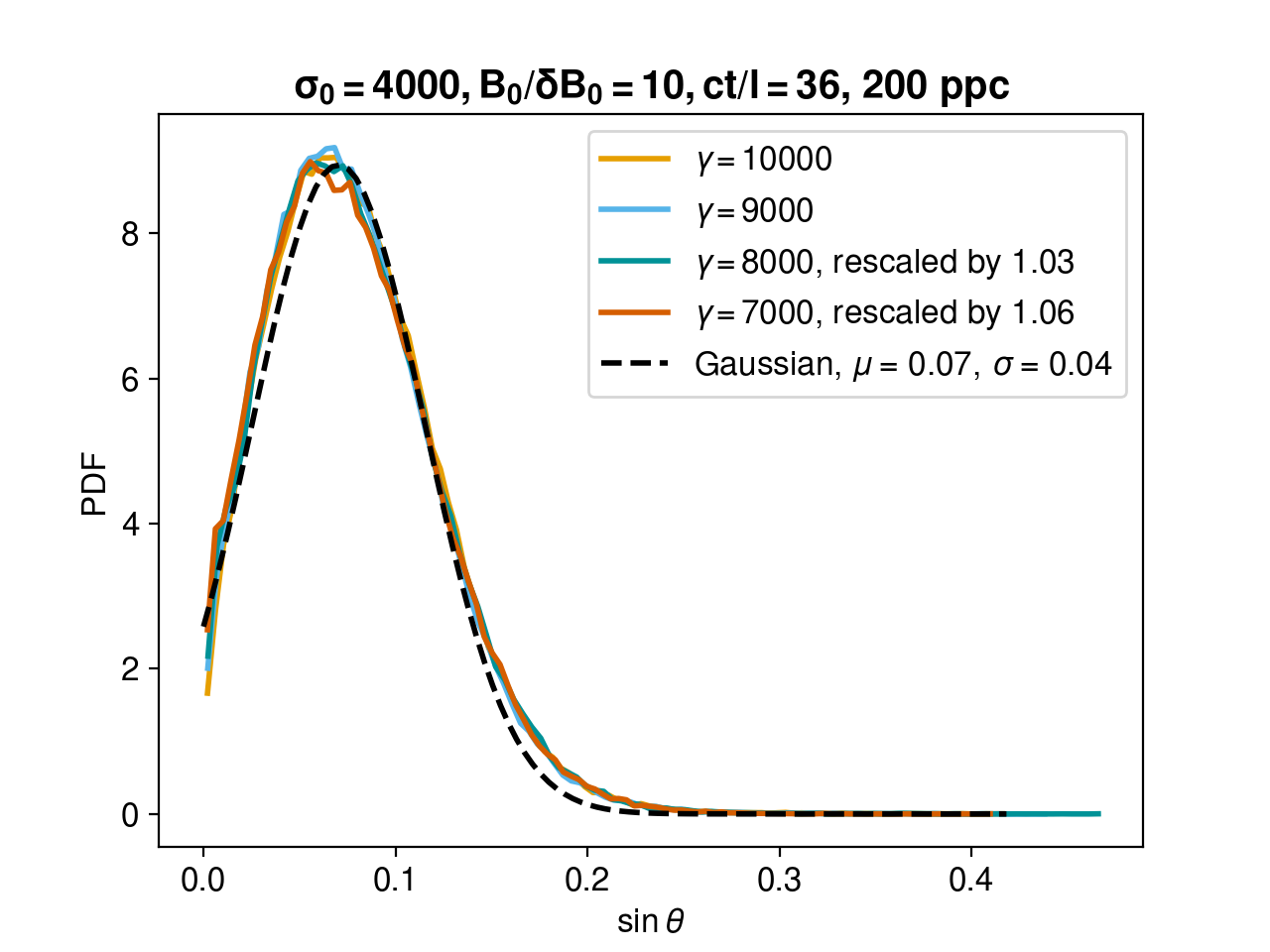}
\caption{  Distribution of pitch angles of tracer particles measured after integrating particle trajectories in static magnetic fields for $c\Delta t/l=40$, with $\gamma$ from $7000$ to $10000$. In order to better resolve the distribution functions, we used $2\times 10^5$ tracer particles for each value of~$\gamma$. The curves are {empirically} rescaled to show self-similarity, and the distribution is {empirically} fit to a Gaussian. 
\label{g10k_dist}}
\end{figure}

Once~$\gamma$ approaches large values, around $7000$ to $10000$, the pitch-angle distribution becomes steeper than a simple exponential and approaches a Gaussian (Figure~\ref{g10k_dist}), albeit with a slight skew to the right. The pitch angle distribution becomes increasingly independent of $\gamma$. Consequently the rescaling symmetry requires correspondingly small multiplicative factors. The distribution at $\gamma = 9000$ does not require any rescaling to visually match the distribution at $\gamma = 10000$, suggesting that a saturated state has been reached. 

Since a particle's zeroth-order trajectory is along the magnetic field, we conjecture that the upper bound on the average pitch angle is $\sin{\theta} \sim \delta B_0/B_0$. That is, a particle accelerated along the magnetic field to very large $\gamma$ will have sufficiently great inertia to be dynamically unaffected by local magnetic fluctuations, suggesting that its average pitch angle can be estimated as the relative magnitude of those fluctuations experienced by the particle. This is consistent with our pitch-angles at very high $\gamma$ saturating close to $\delta B_0/ B_0 \sim 0.1$ in Figure~\ref{sqrt_fit}, and also with the discussion in \cite[][]{vega2024b}. 

\begin{figure}[h!]
\centering
\includegraphics[width=1.05\columnwidth]{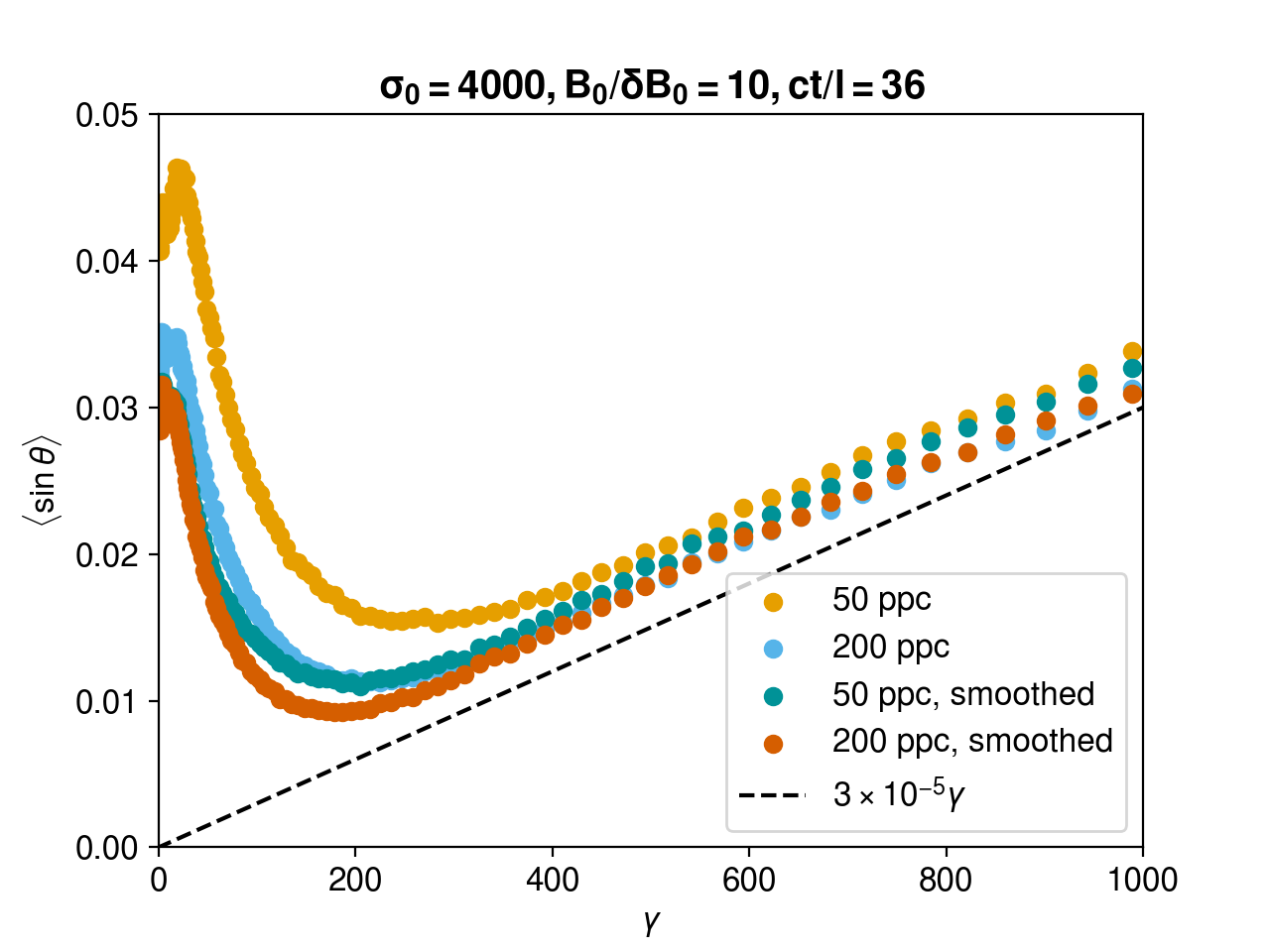}
\caption{ Pitch angles of tracer particles measured after integrating particle trajectories in static magnetic fields for $c\Delta t/l=40$ with linear interpolation. Each dot corresponds to a pitch angle value averaged over $5,000$ tracer particles. Compared to Figure~\ref{noise_comparison}, the pitch-angle broadening is even more affected by the numerical noise at $\gamma\lesssim 300$, though for larger $\gamma$ the slope converges to match the case of bilinear interpolation. 
\label{noise_comparison_yee}}
\end{figure}

Finally, we point out another possible source of numerical error, which is related to the interpolation procedure used to reconstruct the magnetic field inside numerical cells. Compared to the interpolation method used in our study, the VPIC code uses a less smooth (energy conserving) interpolation scheme where magnetic field is interpolated from faces of a staggered Yee mesh~\cite[][]{bowers2008, yee1966}:
\begin{equation}
\begin{aligned}
B_x &= B_x(x_i, y_{j+1/2},z_{k+1/2}), \\ B_y &= B_y(x_{i+1/2}, y_j,z_{k+1/2}), \\ B_z &= B_z(x_{i+1/2}, y_{j+1/2},z_k).
\end{aligned}
\end{equation}
Only the two corresponding faces (e.g., $B_x(x_i, \dots)$ and $B_x(x_{i+1}, \dots)$) are used to interpolate within the cell. In 2D, the $B_z$ component reduces further to a nearest-neighbor interpolation, as there is only one cell face with $\hat{z}$~normal. 
When we repeated our test-particle simulations using this  interpolation scheme, we found even greater noise issues for $\gamma \lesssim 300 $ (Figure~\ref{noise_comparison_yee}), though increasing the particles per cell and smoothing the fields still helped reduce the effects of noise. For larger~$\gamma$, the two interpolation schemes converge to give indistinguishable results. 

We find order-of-magnitude agreement for particle trajectory integration times comparable to the full PIC simulation runtime. However, if tracing in a stationary field is continued indefinitely, particles will pitch-angle scatter slowly and without bound, breaching our estimated $0.1$~threshold after several hundred $c\Delta t/l$ of simulation time. The noisier field of Run I may be slightly more susceptible to this effect than Run II, as the pitch angles can be seen to be somewhat uniformly higher for all~$\gamma$, regardless of smoothing (Figures~\ref{noise_comparison} and~\ref{sqrt_fit}). We conjecture that this may also be an artifact of the interpolation procedure. When the transitions of the fields between cells are not smooth enough, the analytical assumptions regarding the curvature of the magnetic field lines may be violated, consequently affecting the evolution of the adiabatic invariant and the pitch angle. {A similar scenario is analytically studied in \cite[]{kulsrud1957}, wherein discontinuity in the \textit{N}th derivative of the frequency $\omega(t)$ of a time-dependent harmonic oscillator causes the associated adiabatic invariant to be preserved only up to \textit{N}th order in the rate of change of~$\omega$.} {Though $C^1$ continuity in the magnetic field interpolation scheme would be sufficient to keep the curvature finite, discontinuity in the \textit{N}th derivative could still potentially affect particles with sufficiently small angular collimation with the magnetic field.}

\section{Discussion and Conclusion}
We have investigated the acceleration of relativistic particles in magnetically dominated Alfvénic turbulence, specifically when a strong magnetic guide field is present in the plasma. In this scenario, the conservation of the first adiabatic invariant becomes crucial, as it imposes significant restrictions on the evolution of the particle's pitch angle. This conservation law alters the acceleration process considerably, has important implications for interpreting the synchrotron radiation produced by energetic electrons, and imposes nontrivial constraints on numerical studies of turbulent particle acceleration.  

In a pair plasma, when the energy of the initial magnetic fluctuations exceeds the rest-mass energy of plasma particles, the acceleration process occurs in three stages. First, the particle's gyroradius is preserved, and acceleration takes place along the magnetic field lines. During this stage, the particle's pitch angle progressively decreases {to very small values, until being constrained by the intrinsic curvature of the magnetic field}. For instance, in our case of a moderately strong guide field, $B_0/\delta B_0 = 10$, the pitch angle reduces to approximately $10^{-2}$. The corresponding pitch-angle distribution function scales in a self-similar manner with the particle's energy.

In the second stage, as the particle's energy increases, the pitch angle begins to rise due to progressively increasing drifts associated with local magnetic field curvature and gradients. When the particle's gyroradius becomes comparable to the inner scale of Alfv\'enic turbulence,~$d_{rel}$, the acceleration process transitions into the third stage, where the pitch angle keeps increasing at a slower rate and eventually saturates at a value around $\sin\theta \sim \delta B_0/B_0$.

{
The discussed evolution of pitch angles has implications for both the physics of synchrotron radiation and the numerical studies of particle acceleration. The traditional treatment of synchrotron radiation often assumes that the distribution of pitch angles is isotropic. It begins with an expression for the power of synchrotron radiation emitted by an electron with energy ~$\gamma$:
\begin{eqnarray}
\label{p1}
P= c\sigma_T\frac{B_0^2}{4\pi}\gamma^2\sin^2\theta,
\end{eqnarray}
where $\sigma_T$ is the Thomson electron cross section. It then takes into account that an electron with energy $\gamma$ radiates at a characteristic frequency $\nu\propto \gamma^2 B_0\sin\theta $, multiplies Eq.~(\ref{p1}) by the distribution function of the electron energies, $f(\gamma)d\gamma$, and averages over the isotropic distribution of pitch angles. 

Let us assume that the energy distribution function has a power-law form, $f(\gamma)\propto \gamma^{-\delta}$, and that the source is optically thin, that is, the radiation quickly escapes the system. As a result, one gets the spectral distribution of the emitted radiation, that is, the power of radiation per frequency interval~$d\nu$ \cite[e.g.,][]{pacholczyk1970}:
\begin{eqnarray}
\label{j_old}
F_\nu d\nu \propto B_0^{(1+\delta)/2}\nu^{(1-\delta)/2}d\nu.
\end{eqnarray} 

This formula allows one to use the observational spectral exponent of synchrotron radiation $F_\nu\propto \nu^{-\alpha}$, to infer the energy spectrum of the radiating electrons, $\delta=2\alpha +1$.  Moreover, the spectral intensity of radiation allows one to evaluate the strength of the magnetic field in the radiating object, its magnetic cooling time, and other characteristics. 

If, instead, one assumes that the pitch angle is not isotropic, but rather obeys the relation given, for example, by our Eq.~(\ref{square_gamma}), an analysis similar to that outlined in  \cite[][]{pacholczyk1970}, leads to a spectrum of synchrotron radiation significantly different from the conventional Eq.~(\ref{j_old}):
\begin{eqnarray}
\label{j_new}
F_\nu d\nu \propto \left(\frac{\delta B_0}{B_0\,l^{1/3}} \right)^{3(1+\delta)/5} B_0^{(1+\delta)/5}\nu^{(3-2\delta)/5}d\nu.\,
\end{eqnarray}
Applying the conventional Eq.~(\ref{j_old}) instead of Eq.~(\ref{j_new}) to sources with anisotropically accelerated electrons could lead to incorrect conclusions about the particle distribution functions and magnetic energies in the observed objects.}

{On the numerical side,} when a strong guiding field is present, the distribution function of accelerated particles decreases rapidly with increasing energy. Consequently, it becomes difficult to generate a substantial statistical ensemble of particles that have been accelerated to high gamma factors. In our PIC simulations, although our measurements during the first stage of acceleration (that is, at $\gamma \lesssim 200$) were reliable, the accuracy of our measurements for the later stages (at $\gamma > 200$) was insufficient. 

To address these numerical challenges and extend our analysis to energies beyond $\gamma \sim 200$, we utilized a different approach: we examined the angular distributions of test particles propagating in a stationary magnetic field derived from our PIC simulations. This method enabled us to analyze particle angular broadening at relatively high energies, reaching up to $\gamma \sim 10^4$. In the studied energy range, we observed a reasonably good agreement with the phenomenological model proposed in \cite{vega2024b,vega2025}.

Finally, our test-particle studies revealed significant numerical limitations specific to cases of a strong guide field. Since the first adiabatic invariant is well preserved, the gyroradius of particles may remain comparable to or even smaller than the numerical cell size for a considerable portion of the acceleration process. We found that this situation can adversely affect angular collimation due to pitch-angle scattering caused by the numerical noise {present} in PIC simulations. Even when the noise is minimal, nonphysical scattering became significant as pitch angles decreased. We demonstrated, however, that increasing the number of particles per cell and filtering out the electromagnetic fluctuations related to numerical noise can help align pitch-angle values more closely with analytical predictions.

Another limitation we observed relates to the interpolation procedure used for numerically reconstructing the magnetic field within a cell. We discovered that even when the magnetic modes associated with numerical noise are filtered out, the choice of interpolation method still significantly impacts pitch angle evolution. A smoother interpolated magnetic field yields better agreement with analytical predictions. We hypothesize that if the transitions of the magnetic fields between cells are not smooth enough, the analytical assumptions regarding the curvature of the magnetic field lines, necessary for preserving the first adiabatic invariant, may be violated, consequently affecting the evolution of the pitch angle.

We believe that our analysis of the angular distributions of accelerated particles and their scaling with particle energy can offer valuable insights into the processes of relativistic turbulent particle acceleration in strong guide fields. {Our results illustrate the importance of continued, comprehensive study in different parameter regimes, particularly for varied guide fields}. Additionally, {our findings} may aid in interpreting synchrotron spectra from relativistic astrophysical sources.

\begin{acknowledgments}
We are grateful to the anonymous reviewer for constructive suggestions, which helped improve the text. We would also like to thank Ludwig B\"oss, Hayk Hakobyan, Noah Hurst, and Arno Vanthieghem for discussing the results and providing valuable comments. This work was supported by the U.S. Department of Energy, Office of Science, Office of Fusion Energy Sciences under award number DE-SC0024362. The work of DH and SB was also supported by the University of Wisconsin-Madison, Office of the Vice Chancellor for Research with funding from the Wisconsin Alumni Research Foundation. The research of SB was also supported in part by grant NSF PHY-2309135 to the Kavli Institute for Theoretical Physics (KITP). VR was also partly supported by NASA grant 80NSSC21K1692. Computational resources were provided by the Texas Advanced Computing  Center (TACC) at the University of Texas at Austin and by the NASA High-End Computing (HEC) Program through the NASA Advanced Supercomputing (NAS) Division at Ames Research Center. 
This research also used resources of the National Energy Research Scientific Computing Center, a DOE Office of Science User Facility supported by the Office of Science of the U.S. Department of Energy under Contract No. DE-AC02-05CH11231 using NERSC awards FES-ERCAP0028833 and FES-ERCAP0033257. 
\end{acknowledgments}

\newpage

\bibliography{references}{}
\bibliographystyle{aasjournalv7}



\end{document}